\documentclass[%
reprint,
superscriptaddress,
amsmath,amssymb,
aps,
prd,
]{revtex4-1}

\usepackage{graphicx}% Include figure files
\usepackage{dcolumn}% Align table columns on decimal point
\usepackage{bm}% bold math
\usepackage{hyperref}% add hypertext capabilities
\usepackage{subfigure}
\begin{document}

\preprint{APS/123-QED}

\title{New Results on Charged Compact Boson Stars}% Force line breaks with \\
%\thanks{A footnote to the article title}%

\author{Sanjeev Kumar}
\email{sanjeev.kumar.ka@gmail.com}
\affiliation{Department of Physics and Astrophysics, University of Delhi, Delhi-110007, India}
\author{Usha Kulshreshtha}%
\email{ushakulsh@gmail.com, ushakuls@iastate.edu}
\affiliation{Department of Physics, Kirori Mal college, University of Delhi, Delhi-110007, India}%
\affiliation{Department of Physics and Astronomy, Iowa State University, Ames, 50010 IA}
\author{Daya Shankar Kulshreshtha}
\email{dskulsh@gmail.com, dayakuls@iastate.edu}
\affiliation{Department of Physics and Astrophysics, University of Delhi, Delhi-110007, India}%
\affiliation{Department of Physics and Astronomy, Iowa State University, Ames, 50010 IA}

\date{\today}% It is always \today, today,
             %  but any date may be explicitly specified

\begin{abstract}
In this work we present some new results which we have obtained in a study of the phase diagram of charged compact boson stars in the theory involving massive complex scalar fields coupled to the U(1) gauge field and gravity in a conical potential in the presence of a cosmological constant $\Lambda$ which we treat as a free parameter taking positive and negative values and thereby allowing us to study the theory in the de Sitter and Anti de Sitter spaces respectively. In our studies,  we obtain four bifurcation points (possibility of more bifurcation points being not ruled out) in the de Sitter region. We present a detailed discussion of the various regions in our phase diagram with respect to four bifurcation points. Our theory is seen to have rich physics in a particular domain for positive values of $\Lambda$ which is consistent with the accelerated expansion of the universe.
\end{abstract}

\maketitle
%\section{Introduction}
Introduced long ago \cite{Feinblum:1968,Kaup:1968zz,Ruffini:1969qy}, boson stars represent localized self-gravitating solutions studied vary widely in the literature \cite{Jetzer:1991jr,Lee:1991ax,Mielke:2000mh,Liebling:2012fv,Friedberg:1976me,Coleman:1985ki,Kleihaus:2009kr,Kleihaus:2010ep,Hartmann:2012da,Hartmann:2012wa,Hartmann:2013kna,Kumar:2014kna,Kumar:2015sia,Astefanesei:2003qy,Radu:2012yx,Prikas:2004yw,Brihaye:2013hx,Arodz:2008jk,Arodz:2008nm,Arodz:2012zh}. Such theories are being considered in the presence of positive \cite{Hartmann:2013kna,Kumar:2014kna,Kumar:2015sia} as well as negative \cite{Astefanesei:2003qy,Radu:2012yx,Prikas:2004yw,Brihaye:2013hx} values of the cosmological constant $\Lambda$. The theories with positive values of $\Lambda$ (corresponding to the de Sitter (dS) space) are relevant from observational point of view as they describe a more realistic description of the compact stars in the universe since all the observations seem to indicate the existence of a positive cosmological constant. Such theories are also being used to model the dark energy of the universe. However, the theories with negative values of $\Lambda$ (corresponding to the Anti de Sitter (AdS) space) are meaningful in the context of AdS/CFT correspondence \cite{Maldacena:1997re,Witten:1998qj,Brodsky:2011sk}.

In fact, cosmological constant, the value of the energy density of the vacuum of space is the simplest form of dark energy and it  provides a good fit to many cosmological observations. A positive vacuum energy density resulting from a positive cosmological constant (implying a negative) pressure gives an accelerated expansion of the universe consistent with the observations. Our theory is seen to have rich physics in a particular domain for positive values of $\Lambda$.have studied 

In a recent paper \cite{Kumar:2014kna}, we have studied the boson stars and boson shells in a theory of complex scalar field coupled to  $U(1)$ gauge field $A_{\mu}$ and the gravity in the presence of a fixed positive cosmological constant $\Lambda$ (i.e. in the de Sitter space). In the present work we study this theory of complex scalar field coupled to  $U(1)$ gauge field $A_{\mu}$ and the gravity in the presence of a potential: $V(|\Phi|) := (m^2 |\phi|^2 +\lambda |\phi|) $ (with $m$ and $\lambda$ are constant parameters) and a cosmological constant $\Lambda$ which we treat as a free parameter and which takes positive as well as negative values and thereby allowing us to study the theory in the dS as well as in the AdS space.  We investigate the  properties of the solutions of this theory and determine their domains of existence for some specific values of the parameters of the theory. Similar solutions have also been obtained by Kleihaus, Kunz, Laemmerzahl and List, in a V-shaped scalar potential. 

We construct the boson star solutions of this theory numerically and we study their properties. In our studies we investigate in details the phase diagram of the theory for the scalar and the vector fields. In our studies we obtain four bifurcation points (possibility of more bifurcation points being not ruled out) in the dS region. We present a detailed discussion of the various regions in our phase diagram with respect to three bifurcation points.

\noindent We study the theory defined by the action: 
\begin{eqnarray}
S&=&\int \left[ \frac{R-2\Lambda}{16\pi G}   +\mathcal L_M \right] \sqrt{-g}\ d^4\,x\label{3:action}
\nonumber\\ \mathcal L_M& =&	- \frac{1}{4} F^{\mu\nu} F_{\mu\nu}
   -  \left( D_\mu \Phi \right)^* \left( D^\mu \Phi \right)
 - V(|\Phi|)\nonumber\\
 D_\mu \Phi &=& (\partial_\mu \Phi + i e A_\mu \Phi)\ \ ,\ \ F_{\mu\nu} = (\partial_\mu A_\nu - \partial_\nu A_\mu)
%& V(|\Phi|)= m^2|\Phi|^2 +\lambda |\Phi|
\end{eqnarray}
Here $R$ is the Ricci curvature scalar, $G$ is Newton's Gravitational constant and $\Lambda$ is cosmological constant. Also, $g = det(g_{\mu\nu})$ where $g_{\mu\nu}$ is the metric tensor and the asterisk in the above equation denotes complex conjugation. Using the variational principle, equations of motion are obtained as:
\begin{eqnarray}
 G_{\mu\nu}\equiv R_{\mu\nu}-\frac{1}{2}g_{\mu\nu}R = 8\pi G T_{\mu\nu}-\Lambda g_{\mu\nu}
\nonumber \\ 
 \partial_\mu \left ( \sqrt{-g} F^{\mu\nu} \right)%& =&
=   -i\, e \sqrt{-g}\, [\Phi^* (D^\nu \Phi)-\Phi (D^\nu \Phi)^* ] 
\nonumber \\
D_\mu\left(\sqrt{-g}  D^\mu \Phi \right) = 2 m^2 \sqrt{-g}\, \Phi +\frac{\lambda }{2}\sqrt{-g}\,\frac{\Phi}{|\Phi|}\nonumber \\
\left[D_\mu\left(\sqrt{-g}  D^\mu \Phi \right)\right]^* = 2 m^2 \sqrt{-g}\, \Phi^* +\frac{\lambda }{2}\sqrt{-g}\,\frac{\Phi^*}{|\Phi|}
  \label{3:vfeqH}
 \end{eqnarray}
The energy-momentum tensor $T_{\mu\nu}$ is given by ,
\begin{eqnarray}
T_{\mu\nu} &=& \biggl[ ( F_{\mu\alpha} F_{\nu\beta}\ g^{\alpha\beta} -\frac{1}{4} g_{\mu\nu} F_{\alpha\beta} F^{\alpha\beta})
\nonumber\\ & &\quad + (D_\mu \Phi)^* (D_\nu \Phi)+ (D_\mu \Phi) (D_\nu \Phi)^*  
\nonumber \\ & & \quad -g_{\mu\nu} \left((D_\alpha \Phi)^* (D_\beta \Phi)    \right) g^{\alpha\beta}
 -  g_{\mu\nu}\; V( |\Phi|) \biggr]  \ \ \   \label{3:vtmunu}
\end{eqnarray}
To construct spherically symmetric solutions we adopt the static spherically symmetric metric with Schwarzschild like coordinates:
\begin{equation}
ds^2= \biggl[ -A^2 N dt^2 + N^{-1} dr^2 +r^2(d\theta^2 + \sin^2 \theta \;d\phi^2) \biggr]\  \ \
\end{equation}
This leads to the components of Einstein tensor ($G_{\mu\nu}$) 
\begin{eqnarray}
G_t^t &=& \biggl[ \frac{-\left[r\left(1-N\right)\right]'}{r^2} \biggr] ,\  
G_r^r = \biggl[ \frac{2 r A' N -A\left[r\left(1-N\right)\right]'}{A\ r^2} \biggr] \nonumber \\
G_\theta^\theta &=& \biggl[ \frac{2r\left[rA'\ N\right]' + \left[A\ r^2 N'\right]'}{2 A\ r^2} \biggr]
\  \   = \  G_\varphi^\varphi
\end{eqnarray}

Here the arguments of the functions $A(r)$ and $N(r)$ have been suppressed. For solutions with vanishing magnetic field, the Ans\"atze for the matter fields have the form:

\begin{equation}
 \Phi(x^\mu)=\phi(r) e^{i\omega t}\ \ ,\ \ A_\mu(x^\mu) dx^\mu = A_t(r) dt
\end{equation}
We redefine $\phi(r)$ and $A_t(r)$ as:
\begin{equation}
 h(r)={(\sqrt{2} \;e\, \phi(r))}/{m} \ \ \ ,\ \ \  b(r)={(\omega+e A_t(r))}/{m}\label{hb}
 \end{equation}
We introduce new dimensionless constant parameters:
\begin{equation}
 \alpha = \frac{4\pi G\,m^2}{e^2} \ \ \ ,\ \ \  \tilde{\lambda}=\frac{\lambda\,e}{\sqrt{2}\; m^3} \ \ \ ,\ \ \ \tilde{\Lambda}=\frac{\Lambda}{m^2}
\end{equation}
Introducing a dimensionless coordinate $\hat{r}$ defined by $\hat{r}=m\,r$~(implying $\frac{d}{dr}=m\frac{d}{d\hat{r}}$),  Eq. (\ref{hb}) reads as:
\begin{equation}
 h(\hat{r})={(\sqrt{2} \;e\, \phi(\hat{r}))}/{m} \ \ \ ,\ \ \  b(\hat{r})={(\omega+e A_t(\hat{r}))}/{m}\label{hb1}
 \end{equation}
% Defining  ${\rm sign}(h)=0$ if $h=0$ and ${\rm sign}(h)=\pm 1$ for $h>0$ or $h<0$,
 Equations of motion in terms of $h(\hat{r})$ and $b(\hat{r})$ (where the primes denote differentiation with respect to $\hat{r}$ and ${\rm sign }(h)$ denotes the usual signature function) read:
\begin{eqnarray}
\left( A N \hat{r}^2 h'\right)' &=& \frac{\hat{r}^2}{A N}\left[A^2 N(h+\tilde{\lambda}\, {\rm sign}(h)) - b^2 h\right]\label{3:vheq}\\
\left[\big({ \hat{r}^2\  b'}\big)/{A} \right]' &=& \left[ \big({\hat{r}^2 h^2 b }\big)/({A\ N}) \right]\label{3:vbeq}
\end{eqnarray}
We  obtain:
{\begin{widetext}
\begin{subequations}
\begin{eqnarray}
h'' & = &\bigg[ \frac{\alpha\, \hat{r} h'}{A^2N} \left(A^2 h^2 +2 A^2 h \tilde{\lambda} + \,b'^2\right)
-\frac{h'\left(1+N-\tilde{\Lambda} \hat{r}^2\right)}{\hat{r}N}+\frac{A^2 N  h+A^2 N \tilde{\lambda} \, {\rm sign}(h) - b^2 h}{A^2 N^2}\bigg]\label{eq_H}\\
b'' & = & \bigg[\frac{\alpha}{A^2 N^2} \hat{r} b'\left(A^2 N^2 h'^2 + b^2 h^2\right) -\frac{2 b'}{\hat{r}} + \frac{b h^2}{N}\bigg]\ \ ,\ \ \ \ 
A'  = \bigg[ \frac{\alpha \hat{r}}{A N^2}\left(A^2 N^2 h'^2 + b^2 h^2\right) \bigg]\ \\\label{eq_b}
N' & = &\bigg[ \frac{1-N-\tilde{\Lambda} \hat{r}^2}{\hat{r}} -\frac{\alpha \hat{r}}{A^2 N}
\bigg(A^2 N^2 h'^2 + N b'^2 +  b^2 h^2 + A^2 N  h^2 + 2 A^2 N h \tilde{\lambda}\bigg)\bigg] \ 
\label{eq_N}
\end{eqnarray}	
\end{subequations}	
\end{widetext}}
%\newpage
%###################################
%\section{The boundary conditions and global charges}
For the metric function $A(\hat{r})$ we choose the 
boundary condition
%\begin{eqnarray}
$A(\hat{r}_o)=1 \label{Aro} $
%\end{eqnarray}
where $\hat{r}_o$ is the outer radius of the star. For constructing globally regular ball-like boson star solutions, we choose:
\begin{eqnarray}
& N(0)=1 \ ,\ \   b'(0)=0 \ ,\nonumber\\ &h'(0)=0\ ,\ \ h(\hat{r}_o)=0\ , \ \  h'(\hat{r}_o)=0  \label{bcstar}
\end{eqnarray}

For the positive and negative $\tilde{\Lambda}$ we match in the exterior region $\hat{r}>\hat{r}_o$, the Reissner-Nordstr\"om de Sitter and  Reissner-Nordstr\"om Anti de Sitter solutions respectively. The conserved Noether current is given by:
\begin{eqnarray}
j^\mu=-i \,e\,\left[ \Phi(D^\mu \Phi)^*-\Phi^* (D^\mu \Phi) \right]\ \ \ ,\ \  
D_\mu\,j^{\mu} = 0\qquad\
\end{eqnarray}
The charge $Q$ of the boson star is given by
\begin{equation}
Q=-\frac{1}{4\pi}\int_{0} ^{\hat{r}_o} j^t \sqrt{-g} \,dr\,d\theta\,d\phi  \,,\  
j^t=-\frac{h^2(\hat{r}) b(\hat{r})}{A^2(\hat{r}) N(\hat{r})}\quad \nonumber
\end{equation}

For all the gravitating solutions we obtain the mass parameter M (in the units employed):
 \begin{equation}
M= \biggl(1-N(\hat{r}_o)+\frac{\alpha Q^2}{\hat{r}_o^2} -\frac{\tilde{\Lambda}}{3} \hat{r}_o^2\biggr)\frac{\hat{r}_o}{2}
 \end{equation}

\begin{figure}
\begin{center}
	 \mbox{\subfigure[][]{\includegraphics[width=0.96\linewidth,height=0.19\textheight]{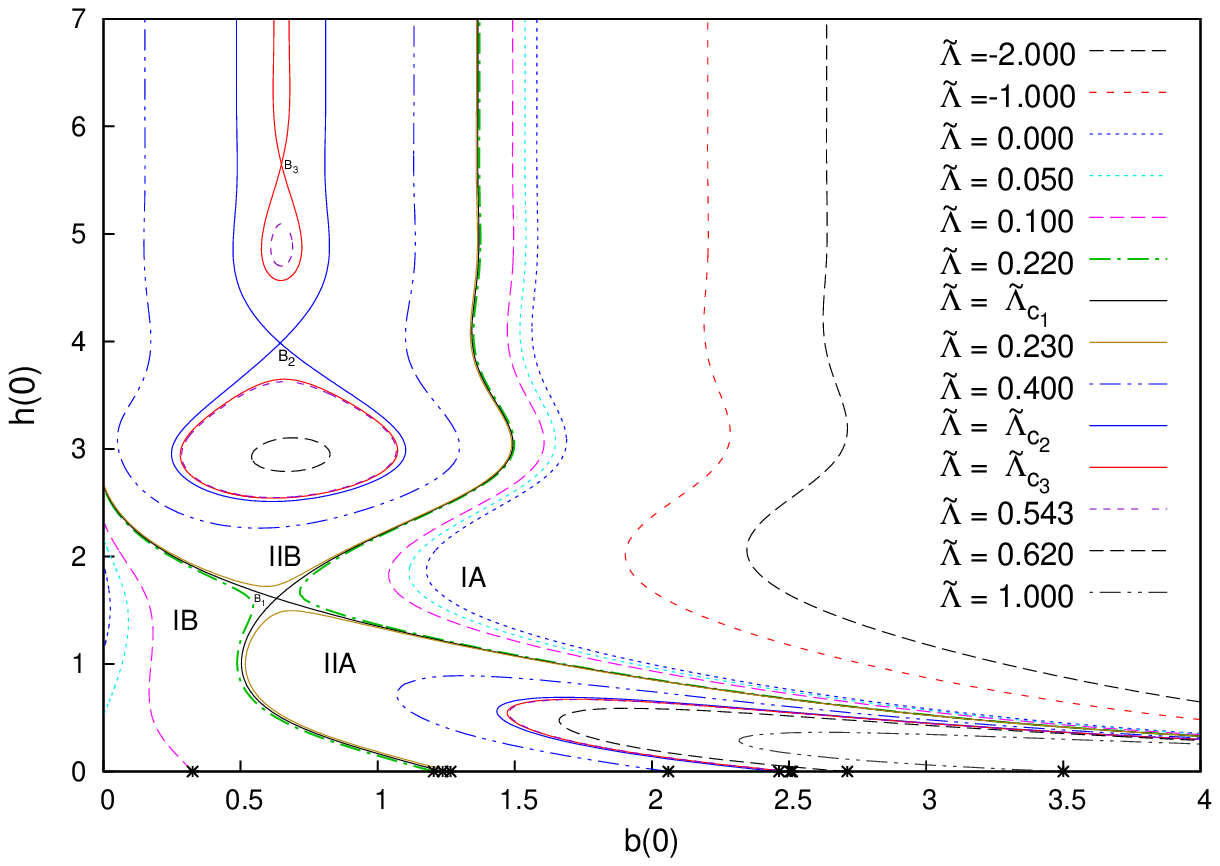}\label{f1a}}}
	\mbox{\subfigure[][]{\includegraphics[width=1.0\linewidth,height=0.19\textheight]{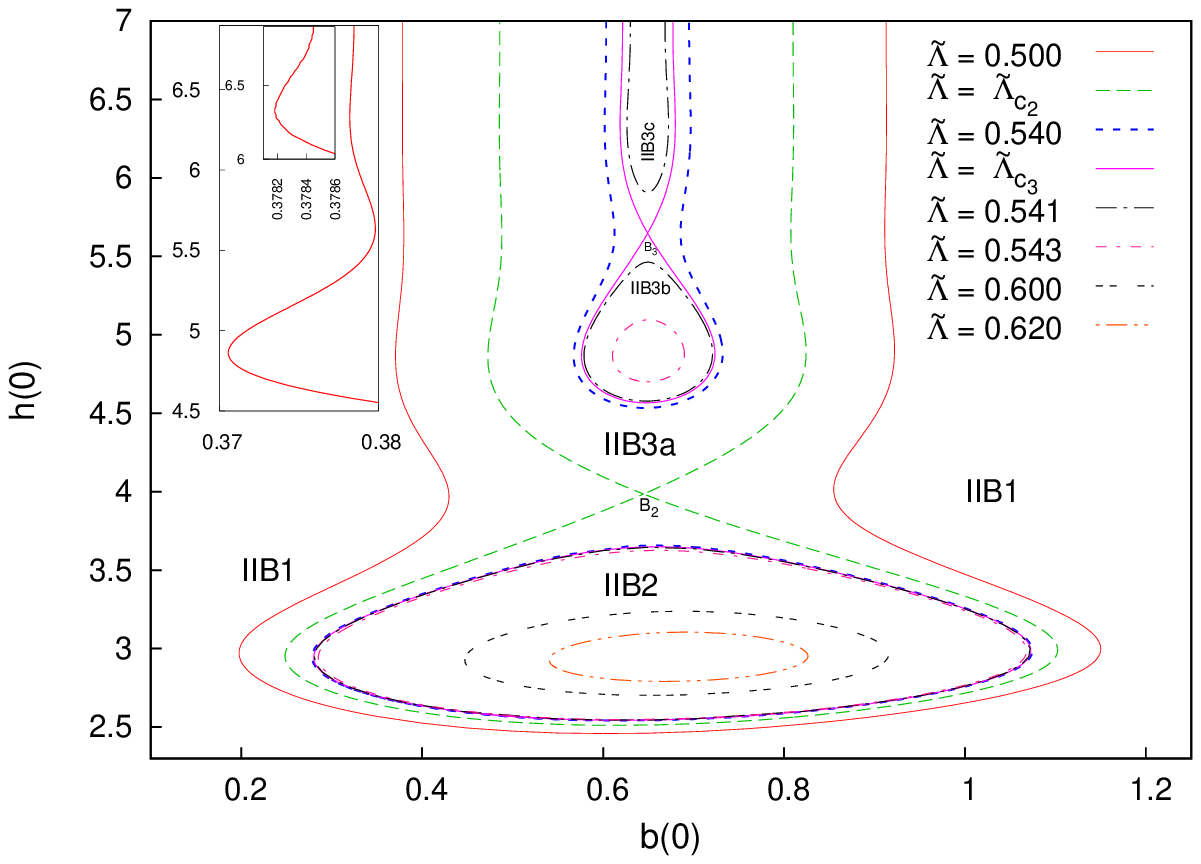}\label{f1b}}}
        \mbox{\subfigure[][]{\includegraphics[width=1.0\linewidth,height=0.189\textheight]{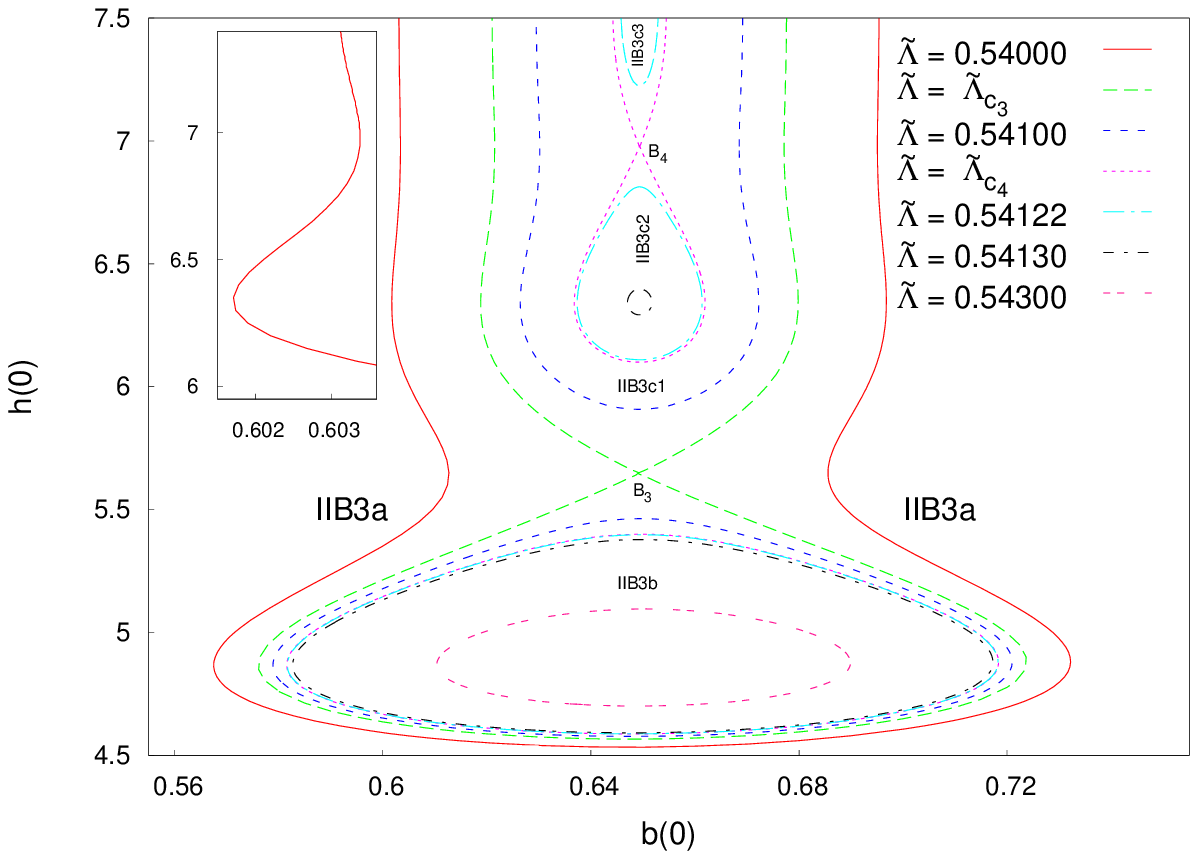}\label{f1c}}}
	\caption{(color online) Fig. (a) depicts the phase diagram of the theory for the vector field  at the center of the star $b(0)$ and the scalar field at the center of the star $h(0)$ for different values of the cosmological constant $\tilde{\Lambda}$ in the range $\tilde{\Lambda}=-2.000$ to $\tilde{\Lambda}=+1.000$. The points $B_1,~B_2 ,~B_3$ and $B_4$ represent the four bifurcation points. The entire region depicted in the phase diagram in Fig. (a) is divided into four regions IA, IB and IIA, IIB in the vicinity of $B_1$. The region IIB of the phase diagram shown in Fig. (a) is separately depicted in details in Fig. (b). The region IIB of the phase diagram is subdivided into three regions IIB1, IIB2 and IIB3 in the vicinity of $B_2$. The region IIB3 is further subdivided into the regions IIB3a, IIB3b and IIB3c in the vicinity of  $B_3$. Similarly the region IIB3c is subdivided in to the regions IIB3c1, IIB3c2 and IIB3c3 in the vicinity of $B_4$. The asterisks shown in Fig. (a), corresponding to $h(0)=0$, represent the transition points from the boson stars to the boson shells. The Figure shown in the inset in Fig. (b) represents a part of the phase diagram with better precision.
\label{fig1}}
\end{center}
\end{figure}
\begin{figure}
\begin{center}
	\mbox{\subfigure[][]{\includegraphics[width=1.0\linewidth,height=0.2\textheight]{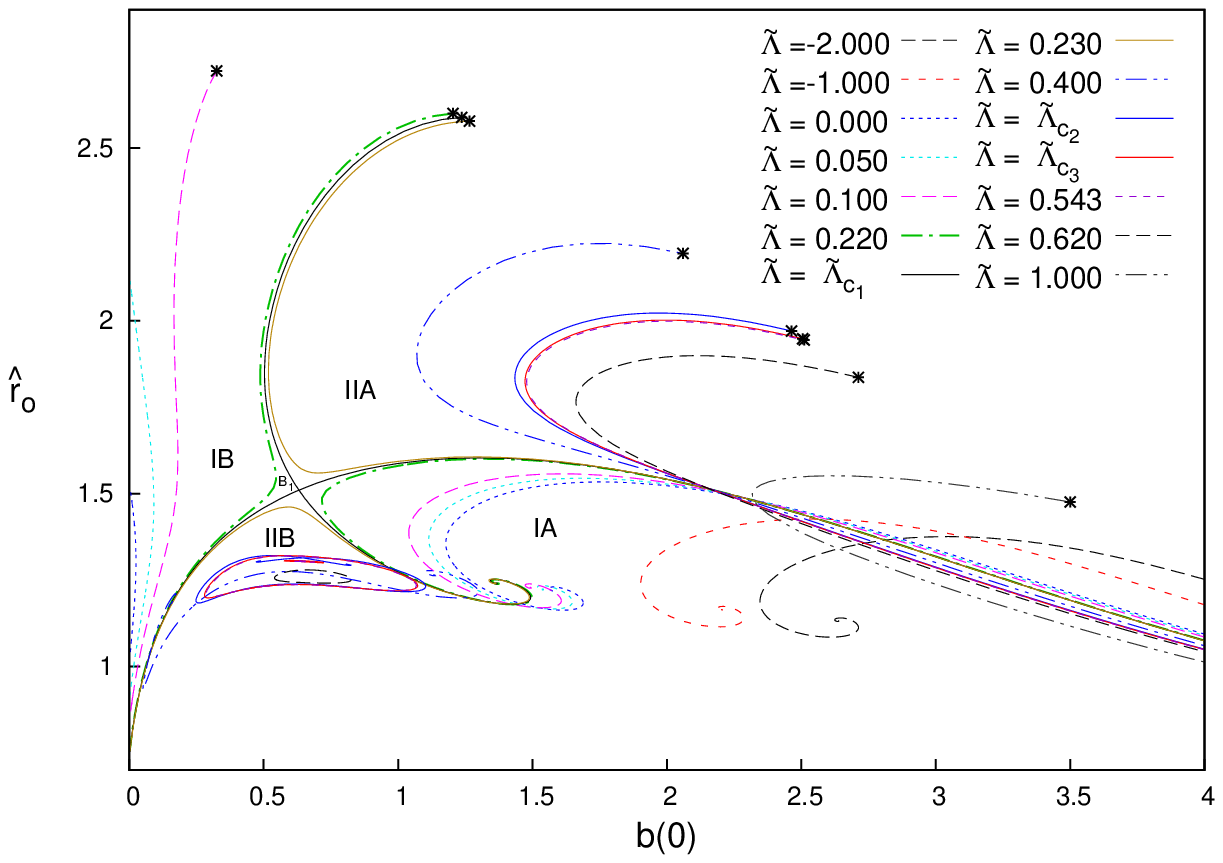}\label{f2a}}}
	\mbox{\subfigure[][]{\includegraphics[width=1.0\linewidth,height=0.2\textheight]{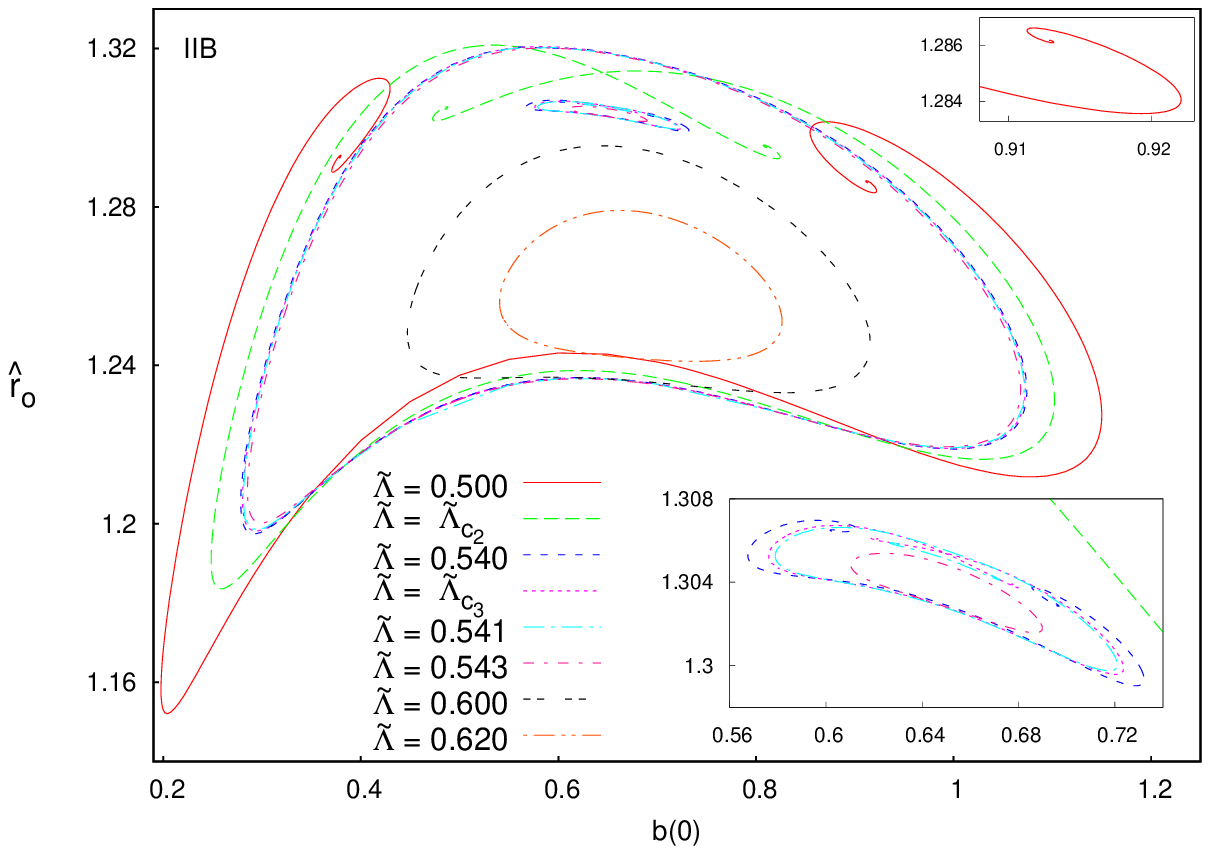}\label{f2b}}}
	\caption{(color online) Fig.(a) shows a plot of the vector field at the center of the star $b(0)$ versus the radius $\hat{r}_o$ of the boson star. The point $B_1$ corresponds to the bifurcation point and the entire region depicted in Fig. (a) is divided into four regions IA, IB and IIA, IIB in the vicinity of $B_1$. The region IIB shown in Fig. (a) is separately depicted in details in Fig. (b). The asterisks shown in Fig. (a) represent the transition points from the boson stars to the boson shells. The spiral behavior of the solutions is visible in the regions IA and IIB. The Figure shown in the inset in Fig. (b) represents a part of the region IIB with better precision. \label{fig2}}
\end{center}
\end{figure}
%\newpage

We now study the numerical solutions of Eqs. (\ref{eq_H})-(\ref{eq_N}) with the boundary conditions defined by  $A(\hat{r}_o)=1$ and Eq. (\ref{bcstar}) and determine their domain of existence for some specific values of the parameters of the theory. Our theory has three parameters $\alpha$, $\tilde{\lambda}$ and $\tilde{\Lambda}$ and we study the theory for different values of $\tilde{\Lambda}$ giving it positive as well as negative values, keeping $\alpha$ and $\tilde{\lambda}$ fixed (namely, $\alpha=0.2$ and $\tilde{\lambda}=1.0$)  and we discuss the corresponding physics as it is observed in our phase diagram. 

We study the phase diagram of the theory involving the vector and scalar fields at the center of the boson star for the different values of the cosmological constant $\tilde{\Lambda}$. We observe some interesting phenomena near some specific values of $\tilde{\Lambda}$ where the system is seen to have bifurcation points $B_1\,,\,B_2\,,\,B_3$ and $B_4$ which correspond to four different values of the cosmological constant $\tilde{\Lambda}$: $\tilde{\Lambda}_{c_1}\simeq0.22521, ~\tilde{\Lambda}_{c_2}\simeq0.52605$ ,  $\tilde{\Lambda}_{c_3}\simeq0.54076 $ and $\tilde{\Lambda}_{c_4}\simeq0.541250 $ respectively (the possibility of some more bifurcation points not being ruled out). 
The theory is seen to have rich physics in the domain $\tilde{\Lambda} = +0.500$ to $\tilde{\Lambda}\simeq + 0.62$.

For a meaningful discussion, we divide our phase diagram into four regions denoted by IA, IB, IIA and IIB in the vicinity of $B_1$ (as seen in Fig. \ref{f1a} ). The asterisks seen in Fig. \ref{f1a} coinciding with the axis $b(0)$ (i.e. corresponding to $h(0)=0$), represent the transition points from the boson stars to boson shells. 

The regions IA, IB and IIA  do not have any bifurcation points, however, the region IIB is seen to contain rich physics evidenced by the occurrence of more bifurcation points in this region. For better details, the region IIB is also plotted in Fig. \ref{f1b} .This region IIB is further divided into the regions IIB1, IIB2 and IIB3 in the vicinity of $B_2$ as seen in Fig. \ref{f1b}. 

The region IIB3 is seen to have further bifurcation point $B_3$. In the vicinity of  $B_3$ we further subdivide the phase diagram into the regions IIB3a, IIB3b and  IIB3c as seen in Fig. \ref{f1b}. The region IIB3b is seen to have closed loops and the behavior of the phase diagram in this region is akin to that of the region IIB2. %On the other hand,  the region IIB3c could in principle contain further bifurcation points and the behaviour of the phase diagram in this region is akin to that of the region IIB1 and IIB3a.
Also the Figure shown in the inset in Figs. \ref{f1b} and  \ref{f1c} represents a part of the phase diagram with better precision.

The region IIB3c is again seen to have further bifurcation point $B_4$, and in the vicinity of $B_4$, we again subdivide the phase diagram in to the regions IIB3c1, IIB3c2 and IIB3c3 as seen in Fig. \ref{f1b} ( and in Fig. \ref{f1b}). The region IIB3c2 is again seen to have closed loops and the behavior of the phase diagram in this region is akin to that of the regions IIB2 and IIB3b. On the other hand, the region IIB3c3 could in principle contain further bifurcation points and the behavior of the phase diagram in this region is akin to that of the region IIB1, IIB3a and IIB3c1.

The regions IA and IB could be divided into two sub-regions corresponding to positive and negative values of $\tilde{\Lambda}$, implying  the dS and AdS regions corresponding to positive and negative values of $\tilde{\Lambda}$. In the region IA, as we change the value of $\tilde{\Lambda}$ in the AdS region from $\tilde{\Lambda}=0.000 $ to $\tilde{\Lambda}=-2.000$ , we observe a continuous deformation of the curves in the phase diagram. In the region IB, as we change the value of $\tilde{\Lambda}$ in the domain $\tilde{\Lambda}=0.000$ to $\tilde{\Lambda}\simeq-0.02$ the theory is seen to have solutions for the boson stars only, without having transition points from boson stars to boson shells and the curves corresponding to the solutions disappear in the phase diagram of the theory for the values $\tilde{\Lambda}\lesssim-0.02$~. 

As we change the value of $\tilde{\Lambda}$ in the dS region from   $\tilde{\Lambda}=0.000 $ to $\tilde{\Lambda}=1.000$ , we observe a lot of new rich physics. While going from  $\tilde{\Lambda}=0.000 $ to some critical value $\tilde{\Lambda}=\tilde{\Lambda}_{c_1}$ , we observe that the solutions exist in two separate domains IA and IB ( as seen in Fig. \ref{f1a}). However, as we increase $\tilde{\Lambda}$ beyond $\tilde{\Lambda} =\tilde{\Lambda}_{c_1}$ the solutions of the theory are seen to exist in the regions IIA and IIB (instead of the regions IA and IB).

As we increase the value of $\tilde{\Lambda}$ from one critical value $\tilde{\Lambda}=\tilde{\Lambda}_{c_1}$ to another critical value $\tilde{\Lambda}=\tilde{\Lambda}_{c_2}$, we notice that the region IIA in the phase diagram shows a continuous deformation of the curves and the region IIB is seen to have its own rich physics as explained in the foregoing. 

As we increase $\tilde{\Lambda}$ beyond $\tilde{\Lambda}_{c_2}$, we observe that in the region IIA there is again a continuous deformation of the curves all the way up to $\tilde{\Lambda}=1.000$. However in the region IIB, we encounter another bifurcation point which divides the region IIB in to IIB1, IIB2 and IIB3. We observe that in the region IIB1 there is a continuous deformation of the curves and the region IIB2 contains closed loops of the curves. The region IIB3 is subdivided into the regions IIB3a, IIB3b and IIB3c. The region IIB3a would have a continuous deformation of the curves and the region IIB3b is seen to contain closed loops. The region IIB3c (subdivided into the regions IIB3c1, IIB3c2 and IIB3c3) has its own rich physics as depicted in Figs. \ref{f1b} and \ref{f1c} as discussed in details in the foregoing.  The region IIB3c3 has its own rich physics in the sense that this region, could in principle, have further bifurcation points.

A plot of the vector field at the center of the star $b(0)$ versus the radius $\hat{r}_o$ of the boson star is depicted in Fig. \ref{f2a}. As before, the point $B_1$ corresponds to the bifurcation point and the entire region depicted in Fig. \ref{f2a} is divided into four regions IA, IB and IIA, IIB in the vicinity of the bifurcation point $B_1$. The region IIB shown in Fig. \ref{f2a} is separately depicted in details in Fig. \ref{f2b}. The asterisks shown in Fig. \ref{f2a} represent the transition points from the boson stars to the boson shells. The spiral behavior of the solutions is visible in the regions IA and IIB. The Figure shown in the inset in Fig. \ref{f2b} represents a part of the region IIB with a better precision.

In conclusion, we have studied in this work a theory of massive complex scalar field coupled to the U(1) gauge field and the gravity with a conical potential in the presence of a cosmological constant $\Lambda$ which takes positive as well as negative values. The theory is seen to have rich physics in the domain $\tilde{\Lambda}=0.5$ to $\tilde{\Lambda}\simeq0.62$. Four bifurcation points $B_1,~B_2,~B_3$ and $B_4$ have been obtained in the phase diagram and the physical behavior of the phase diagram has been  discussed in the various regions of the phase diagram. We have observed interesting physics near the four bifurcation points which correspond to the positive values of $\tilde{\Lambda}$.

Towards the end we make some interesting observations on our studies. Our theory has three free parameters. 
If we fix any two of them at some appropriate values and vary the third carefully then we notice that the bifurcation phenomenon occurs, suggesting that the bifurcation phenomenon seems to be generic. In the present studies we have fixed $\alpha=0.2$ and $\tilde{\lambda}=1.0$ and have studied the theory by varying the value of $\tilde{\Lambda}$ from $-2.0$ to $1.0$ in the phase diagram.

We wish to emphasize that in particular, if we for example, fix $\tilde{\Lambda}=0.541$ , $\alpha=0.2$ and $\tilde{\lambda}=1.0$ then we obtain closed loops in the phase diagram and if we vary any one of the parameters keeing the other two parameters fixed then we obtain bifurcation points between these closed loops as seen in Fig. \ref{f1b} for the variation of $\tilde{\Lambda}$. 

The results of our preliminary investigations suggest that in particular, if we fix  $\tilde{\Lambda} = 0$ and $\alpha = 0.2$ and vary $\tilde{\lambda}$ carefully then we notice the bifurcation phenomenon in an analogous manner with our present studies. Following the same logic, if we fix $\tilde{\Lambda}$ and $\tilde{\lambda}$ appropriately and vary $\alpha$ carefully then we again expect to obtain a bifurcation phenomenon. These investigations are currently in the process and the detailed results would be reported later in a separate communication. Nevertheless we feel that the occurence of the bifurcation phenomenon should be a generic feature of the theory.

\section*{Acknowledgment}
We thank Jutta Kunz and Burkhard Kleihaus for introducing us to this beautiful subject and for helpful educative discussions and encouragements. We also thank James Vary for very useful discussions. This work was supported in part by the US Department of Energy under Grant No. DE-FG02-87ER40371 and by the US National Science Foundation under Grant No. PHY-0904782.
%\nocite{*}
\bibliography{prdr}

%merlin.mbs apsrev4-1.bst 2010-07-25 4.21a (PWD, AO, DPC) hacked
%Control: key (0)
%Control: author (8) initials jnrlst
%Control: editor formatted (1) identically to author
%Control: production of article title (-1) disabled
%Control: page (0) single
%Control: year (1) truncated
%Control: production of eprint (0) enabled
\begin{thebibliography}{26}%
\makeatletter
\providecommand \@ifxundefined [1]{%
 \@ifx{#1\undefined}
}%
\providecommand \@ifnum [1]{%
 \ifnum #1\expandafter \@firstoftwo
 \else \expandafter \@secondoftwo
 \fi
}%
\providecommand \@ifx [1]{%
 \ifx #1\expandafter \@firstoftwo
 \else \expandafter \@secondoftwo
 \fi
}%
\providecommand \natexlab [1]{#1}%
\providecommand \enquote  [1]{``#1''}%
\providecommand \bibnamefont  [1]{#1}%
\providecommand \bibfnamefont [1]{#1}%
\providecommand \citenamefont [1]{#1}%
\providecommand \href@noop [0]{\@secondoftwo}%
\providecommand \href [0]{\begingroup \@sanitize@url \@href}%
\providecommand \@href[1]{\@@startlink{#1}\@@href}%
\providecommand \@@href[1]{\endgroup#1\@@endlink}%
\providecommand \@sanitize@url [0]{\catcode `\\12\catcode `\$12\catcode
  `\&12\catcode `\#12\catcode `\^12\catcode `\_12\catcode `\%12\relax}%
\providecommand \@@startlink[1]{}%
\providecommand \@@endlink[0]{}%
\providecommand \url  [0]{\begingroup\@sanitize@url \@url }%
\providecommand \@url [1]{\endgroup\@href {#1}{\urlprefix }}%
\providecommand \urlprefix  [0]{URL }%
\providecommand \Eprint [0]{\href }%
\providecommand \doibase [0]{http://dx.doi.org/}%
\providecommand \selectlanguage [0]{\@gobble}%
\providecommand \bibinfo  [0]{\@secondoftwo}%
\providecommand \bibfield  [0]{\@secondoftwo}%
\providecommand \translation [1]{[#1]}%
\providecommand \BibitemOpen [0]{}%
\providecommand \bibitemStop [0]{}%
\providecommand \bibitemNoStop [0]{.\EOS\space}%
\providecommand \EOS [0]{\spacefactor3000\relax}%
\providecommand \BibitemShut  [1]{\csname bibitem#1\endcsname}%
\let\auto@bib@innerbib\@empty
%</preamble>
\bibitem [{\citenamefont {Feinblum}\ and\ \citenamefont
  {McKinley}(1968)}]{Feinblum:1968}%
  \BibitemOpen
  \bibfield  {author} {\bibinfo {author} {\bibfnamefont {D.~A.}\ \bibnamefont
  {Feinblum}}\ and\ \bibinfo {author} {\bibfnamefont {W.~A.}\ \bibnamefont
  {McKinley}},\ }\href {\doibase 10.1103/PhysRev.168.1445} {\bibfield
  {journal} {\bibinfo  {journal} {Phys. Rev.}\ }\textbf {\bibinfo {volume}
  {168}},\ \bibinfo {pages} {1445} (\bibinfo {year} {1968})}\BibitemShut
  {NoStop}%
\bibitem [{\citenamefont {Kaup}(1968)}]{Kaup:1968zz}%
  \BibitemOpen
  \bibfield  {author} {\bibinfo {author} {\bibfnamefont {D.~J.}\ \bibnamefont
  {Kaup}},\ }\href {\doibase 10.1103/PhysRev.172.1331} {\bibfield  {journal}
  {\bibinfo  {journal} {Phys. Rev.}\ }\textbf {\bibinfo {volume} {172}},\
  \bibinfo {pages} {1331} (\bibinfo {year} {1968})}\BibitemShut {NoStop}%
%%CITATION = PHRVA,172,1331;%%
\bibitem [{\citenamefont {Ruffini}\ and\ \citenamefont
  {Bonazzola}(1969)}]{Ruffini:1969qy}%
  \BibitemOpen
  \bibfield  {author} {\bibinfo {author} {\bibfnamefont {R.}~\bibnamefont
  {Ruffini}}\ and\ \bibinfo {author} {\bibfnamefont {S.}~\bibnamefont
  {Bonazzola}},\ }\href {\doibase 10.1103/PhysRev.187.1767} {\bibfield
  {journal} {\bibinfo  {journal} {Phys. Rev.}\ }\textbf {\bibinfo {volume}
  {187}},\ \bibinfo {pages} {1767} (\bibinfo {year} {1969})}\BibitemShut
  {NoStop}%
%%CITATION = PHRVA,187,1767;%%
\bibitem [{\citenamefont {Jetzer}(1992)}]{Jetzer:1991jr}%
  \BibitemOpen
  \bibfield  {author} {\bibinfo {author} {\bibfnamefont {P.}~\bibnamefont
  {Jetzer}},\ }\href {\doibase 10.1016/0370-1573(92)90123-H} {\bibfield
  {journal} {\bibinfo  {journal} {Phys.Rept.}\ }\textbf {\bibinfo {volume}
  {220}},\ \bibinfo {pages} {163} (\bibinfo {year} {1992})}\BibitemShut
  {NoStop}%
%%CITATION = PRPLC,220,163;%%
\bibitem [{\citenamefont {Lee}\ and\ \citenamefont {Pang}(1992)}]{Lee:1991ax}%
  \BibitemOpen
  \bibfield  {author} {\bibinfo {author} {\bibfnamefont {T.}~\bibnamefont
  {Lee}}\ and\ \bibinfo {author} {\bibfnamefont {Y.}~\bibnamefont {Pang}},\
  }\href {\doibase 10.1016/0370-1573(92)90064-7} {\bibfield  {journal}
  {\bibinfo  {journal} {Phys.Rept.}\ }\textbf {\bibinfo {volume} {221}},\
  \bibinfo {pages} {251} (\bibinfo {year} {1992})}\BibitemShut {NoStop}%
%%CITATION = PRPLC,221,251;%%
\bibitem [{\citenamefont {Mielke}\ and\ \citenamefont
  {Schunck}(2000)}]{Mielke:2000mh}%
  \BibitemOpen
  \bibfield  {author} {\bibinfo {author} {\bibfnamefont {E.~W.}\ \bibnamefont
  {Mielke}}\ and\ \bibinfo {author} {\bibfnamefont {F.~E.}\ \bibnamefont
  {Schunck}},\ }\href {\doibase 10.1016/S0550-3213(99)00492-7} {\bibfield
  {journal} {\bibinfo  {journal} {Nucl.Phys.}\ }\textbf {\bibinfo {volume}
  {B564}},\ \bibinfo {pages} {185} (\bibinfo {year} {2000})},\ \Eprint
  {http://arxiv.org/abs/gr-qc/0001061} {arXiv:gr-qc/0001061 [gr-qc]}
  \BibitemShut {NoStop}%
%%CITATION = GR-QC/0001061;%%
\bibitem [{\citenamefont {Liebling}\ and\ \citenamefont
  {Palenzuela}(2012)}]{Liebling:2012fv}%
  \BibitemOpen
  \bibfield  {author} {\bibinfo {author} {\bibfnamefont {S.~L.}\ \bibnamefont
  {Liebling}}\ and\ \bibinfo {author} {\bibfnamefont {C.}~\bibnamefont
  {Palenzuela}},\ }\href {\doibase 10.12942/lrr-2012-6} {\bibfield  {journal}
  {\bibinfo  {journal} {Living Rev. Rel.}\ }\textbf {\bibinfo {volume} {15}},\
  \bibinfo {pages} {6} (\bibinfo {year} {2012})},\ \Eprint
  {http://arxiv.org/abs/1202.5809} {arXiv:1202.5809 [gr-qc]} \BibitemShut
  {NoStop}%
%%CITATION = ARXIV:1202.5809;%%
\bibitem [{\citenamefont {Friedberg}\ \emph {et~al.}(1976)\citenamefont
  {Friedberg}, \citenamefont {Lee},\ and\ \citenamefont
  {Sirlin}}]{Friedberg:1976me}%
  \BibitemOpen
  \bibfield  {author} {\bibinfo {author} {\bibfnamefont {R.}~\bibnamefont
  {Friedberg}}, \bibinfo {author} {\bibfnamefont {T.}~\bibnamefont {Lee}}, \
  and\ \bibinfo {author} {\bibfnamefont {A.}~\bibnamefont {Sirlin}},\ }\href
  {\doibase 10.1103/PhysRevD.13.2739} {\bibfield  {journal} {\bibinfo
  {journal} {Phys.Rev.}\ }\textbf {\bibinfo {volume} {D13}},\ \bibinfo {pages}
  {2739} (\bibinfo {year} {1976})}\BibitemShut {NoStop}%
%%CITATION = PHRVA,D13,2739;%%
\bibitem [{\citenamefont {Coleman}(1985)}]{Coleman:1985ki}%
  \BibitemOpen
  \bibfield  {author} {\bibinfo {author} {\bibfnamefont {S.~R.}\ \bibnamefont
  {Coleman}},\ }\href {\doibase 10.1016/0550-3213(85)90286-X} {\bibfield
  {journal} {\bibinfo  {journal} {Nucl.Phys.}\ }\textbf {\bibinfo {volume}
  {B262}},\ \bibinfo {pages} {263} (\bibinfo {year} {1985})}\BibitemShut
  {NoStop}%
%%CITATION = NUPHA,B262,263;%%
\bibitem [{\citenamefont {Kleihaus}\ \emph {et~al.}(2009)\citenamefont
  {Kleihaus}, \citenamefont {Kunz}, \citenamefont {Lammerzahl},\ and\
  \citenamefont {List}}]{Kleihaus:2009kr}%
  \BibitemOpen
  \bibfield  {author} {\bibinfo {author} {\bibfnamefont {B.}~\bibnamefont
  {Kleihaus}}, \bibinfo {author} {\bibfnamefont {J.}~\bibnamefont {Kunz}},
  \bibinfo {author} {\bibfnamefont {C.}~\bibnamefont {Lammerzahl}}, \ and\
  \bibinfo {author} {\bibfnamefont {M.}~\bibnamefont {List}},\ }\href {\doibase
  10.1016/j.physletb.2009.03.066} {\bibfield  {journal} {\bibinfo  {journal}
  {Phys.Lett.}\ }\textbf {\bibinfo {volume} {B675}},\ \bibinfo {pages} {102}
  (\bibinfo {year} {2009})},\ \Eprint {http://arxiv.org/abs/0902.4799}
  {arXiv:0902.4799 [gr-qc]} \BibitemShut {NoStop}%
%%CITATION = ARXIV:0902.4799;%%
\bibitem [{\citenamefont {Kleihaus}\ \emph {et~al.}(2010)\citenamefont
  {Kleihaus}, \citenamefont {Kunz}, \citenamefont {Lammerzahl},\ and\
  \citenamefont {List}}]{Kleihaus:2010ep}%
  \BibitemOpen
  \bibfield  {author} {\bibinfo {author} {\bibfnamefont {B.}~\bibnamefont
  {Kleihaus}}, \bibinfo {author} {\bibfnamefont {J.}~\bibnamefont {Kunz}},
  \bibinfo {author} {\bibfnamefont {C.}~\bibnamefont {Lammerzahl}}, \ and\
  \bibinfo {author} {\bibfnamefont {M.}~\bibnamefont {List}},\ }\href {\doibase
  10.1103/PhysRevD.82.104050} {\bibfield  {journal} {\bibinfo  {journal}
  {Phys.Rev.}\ }\textbf {\bibinfo {volume} {D82}},\ \bibinfo {pages} {104050}
  (\bibinfo {year} {2010})},\ \Eprint {http://arxiv.org/abs/1007.1630}
  {arXiv:1007.1630 [gr-qc]} \BibitemShut {NoStop}%
%%CITATION = ARXIV:1007.1630;%%
\bibitem [{\citenamefont {Hartmann}\ \emph {et~al.}(2012)\citenamefont
  {Hartmann}, \citenamefont {Kleihaus}, \citenamefont {Kunz},\ and\
  \citenamefont {Schaffer}}]{Hartmann:2012da}%
  \BibitemOpen
  \bibfield  {author} {\bibinfo {author} {\bibfnamefont {B.}~\bibnamefont
  {Hartmann}}, \bibinfo {author} {\bibfnamefont {B.}~\bibnamefont {Kleihaus}},
  \bibinfo {author} {\bibfnamefont {J.}~\bibnamefont {Kunz}}, \ and\ \bibinfo
  {author} {\bibfnamefont {I.}~\bibnamefont {Schaffer}},\ }\href {\doibase
  10.1016/j.physletb.2012.06.067} {\bibfield  {journal} {\bibinfo  {journal}
  {Phys.Lett.}\ }\textbf {\bibinfo {volume} {B714}},\ \bibinfo {pages} {120}
  (\bibinfo {year} {2012})},\ \Eprint {http://arxiv.org/abs/1205.0899}
  {arXiv:1205.0899 [gr-qc]} \BibitemShut {NoStop}%
%%CITATION = ARXIV:1205.0899;%%
\bibitem [{\citenamefont {Hartmann}\ and\ \citenamefont
  {Riedel}(2012)}]{Hartmann:2012wa}%
  \BibitemOpen
  \bibfield  {author} {\bibinfo {author} {\bibfnamefont {B.}~\bibnamefont
  {Hartmann}}\ and\ \bibinfo {author} {\bibfnamefont {J.}~\bibnamefont
  {Riedel}},\ }\href {\doibase 10.1103/PhysRevD.86.104008} {\bibfield
  {journal} {\bibinfo  {journal} {Phys. Rev.}\ }\textbf {\bibinfo {volume}
  {D86}},\ \bibinfo {pages} {104008} (\bibinfo {year} {2012})},\ \Eprint
  {http://arxiv.org/abs/1204.6239} {arXiv:1204.6239 [hep-th]} \BibitemShut
  {NoStop}%
%%CITATION = ARXIV:1204.6239;%%
\bibitem [{\citenamefont {Hartmann}\ \emph {et~al.}(2013)\citenamefont
  {Hartmann}, \citenamefont {Kleihaus}, \citenamefont {Kunz},\ and\
  \citenamefont {Schaffer}}]{Hartmann:2013kna}%
  \BibitemOpen
  \bibfield  {author} {\bibinfo {author} {\bibfnamefont {B.}~\bibnamefont
  {Hartmann}}, \bibinfo {author} {\bibfnamefont {B.}~\bibnamefont {Kleihaus}},
  \bibinfo {author} {\bibfnamefont {J.}~\bibnamefont {Kunz}}, \ and\ \bibinfo
  {author} {\bibfnamefont {I.}~\bibnamefont {Schaffer}},\ }\href {\doibase
  10.1103/PhysRevD.88.124033} {\bibfield  {journal} {\bibinfo  {journal} {Phys.
  Rev.}\ }\textbf {\bibinfo {volume} {D88}},\ \bibinfo {pages} {124033}
  (\bibinfo {year} {2013})},\ \Eprint {http://arxiv.org/abs/1310.3632}
  {arXiv:1310.3632 [gr-qc]} \BibitemShut {NoStop}%
%%CITATION = ARXIV:1310.3632;%%
\bibitem [{\citenamefont {Kumar}\ \emph {et~al.}(2014)\citenamefont {Kumar},
  \citenamefont {Kulshreshtha},\ and\ \citenamefont
  {Shankar~Kulshreshtha}}]{Kumar:2014kna}%
  \BibitemOpen
  \bibfield  {author} {\bibinfo {author} {\bibfnamefont {S.}~\bibnamefont
  {Kumar}}, \bibinfo {author} {\bibfnamefont {U.}~\bibnamefont {Kulshreshtha}},
  \ and\ \bibinfo {author} {\bibfnamefont {D.}~\bibnamefont
  {Shankar~Kulshreshtha}},\ }\href {\doibase 10.1088/0264-9381/31/16/167001}
  {\bibfield  {journal} {\bibinfo  {journal} {Class.Quant.Grav.}\ }\textbf
  {\bibinfo {volume} {31}},\ \bibinfo {pages} {167001} (\bibinfo {year}
  {2014})}\BibitemShut {NoStop}%
%%CITATION = CQGRD,31,167001;%%
\bibitem [{\citenamefont {Kumar}\ \emph {et~al.}(2015)\citenamefont {Kumar},
  \citenamefont {Kulshreshtha},\ and\ \citenamefont
  {Kulshreshtha}}]{Kumar:2015sia}%
  \BibitemOpen
  \bibfield  {author} {\bibinfo {author} {\bibfnamefont {S.}~\bibnamefont
  {Kumar}}, \bibinfo {author} {\bibfnamefont {U.}~\bibnamefont {Kulshreshtha}},
  \ and\ \bibinfo {author} {\bibfnamefont {D.~S.}\ \bibnamefont
  {Kulshreshtha}},\ }\href {\doibase 10.1007/s10714-015-1918-0} {\bibfield
  {journal} {\bibinfo  {journal} {Gen. Rel. Grav.}\ }\textbf {\bibinfo {volume}
  {47}},\ \bibinfo {pages} {76} (\bibinfo {year} {2015})}\BibitemShut {NoStop}%
%%CITATION = GRGVA,47,76;%%
\bibitem [{\citenamefont {Astefanesei}\ and\ \citenamefont
  {Radu}(2003)}]{Astefanesei:2003qy}%
  \BibitemOpen
  \bibfield  {author} {\bibinfo {author} {\bibfnamefont {D.}~\bibnamefont
  {Astefanesei}}\ and\ \bibinfo {author} {\bibfnamefont {E.}~\bibnamefont
  {Radu}},\ }\href {\doibase 10.1016/S0550-3213(03)00482-6} {\bibfield
  {journal} {\bibinfo  {journal} {Nucl.Phys.}\ }\textbf {\bibinfo {volume}
  {B665}},\ \bibinfo {pages} {594} (\bibinfo {year} {2003})},\ \Eprint
  {http://arxiv.org/abs/gr-qc/0309131} {arXiv:gr-qc/0309131 [gr-qc]}
  \BibitemShut {NoStop}%
%%CITATION = GR-QC/0309131;%%
\bibitem [{\citenamefont {Radu}\ and\ \citenamefont
  {Subagyo}(2012)}]{Radu:2012yx}%
  \BibitemOpen
  \bibfield  {author} {\bibinfo {author} {\bibfnamefont {E.}~\bibnamefont
  {Radu}}\ and\ \bibinfo {author} {\bibfnamefont {B.}~\bibnamefont {Subagyo}},\
  }\href {\doibase 10.1016/j.physletb.2012.09.050} {\bibfield  {journal}
  {\bibinfo  {journal} {Phys.Lett.}\ }\textbf {\bibinfo {volume} {B717}},\
  \bibinfo {pages} {450} (\bibinfo {year} {2012})},\ \Eprint
  {http://arxiv.org/abs/1207.3715} {arXiv:1207.3715 [gr-qc]} \BibitemShut
  {NoStop}%
%%CITATION = ARXIV:1207.3715;%%
\bibitem [{\citenamefont {Prikas}(2004)}]{Prikas:2004yw}%
  \BibitemOpen
  \bibfield  {author} {\bibinfo {author} {\bibfnamefont {A.}~\bibnamefont
  {Prikas}},\ }\href {\doibase 10.1023/B:GERG.0000035955.07614.0d} {\bibfield
  {journal} {\bibinfo  {journal} {Gen.Rel.Grav.}\ }\textbf {\bibinfo {volume}
  {36}},\ \bibinfo {pages} {1841} (\bibinfo {year} {2004})},\ \Eprint
  {http://arxiv.org/abs/hep-th/0403019} {arXiv:hep-th/0403019 [hep-th]}
  \BibitemShut {NoStop}%
%%CITATION = HEP-TH/0403019;%%
\bibitem [{\citenamefont {Brihaye}\ \emph {et~al.}(2013)\citenamefont
  {Brihaye}, \citenamefont {Hartmann},\ and\ \citenamefont
  {Tojiev}}]{Brihaye:2013hx}%
  \BibitemOpen
  \bibfield  {author} {\bibinfo {author} {\bibfnamefont {Y.}~\bibnamefont
  {Brihaye}}, \bibinfo {author} {\bibfnamefont {B.}~\bibnamefont {Hartmann}}, \
  and\ \bibinfo {author} {\bibfnamefont {S.}~\bibnamefont {Tojiev}},\ }\href
  {\doibase 10.1088/0264-9381/30/11/115009} {\bibfield  {journal} {\bibinfo
  {journal} {Class. Quant. Grav.}\ }\textbf {\bibinfo {volume} {30}},\ \bibinfo
  {pages} {115009} (\bibinfo {year} {2013})},\ \Eprint
  {http://arxiv.org/abs/1301.2452} {arXiv:1301.2452 [hep-th]} \BibitemShut
  {NoStop}%
%%CITATION = ARXIV:1301.2452;%%
\bibitem [{\citenamefont {Arodz}\ and\ \citenamefont
  {Lis}(2008)}]{Arodz:2008jk}%
  \BibitemOpen
  \bibfield  {author} {\bibinfo {author} {\bibfnamefont {H.}~\bibnamefont
  {Arodz}}\ and\ \bibinfo {author} {\bibfnamefont {J.}~\bibnamefont {Lis}},\
  }\href {\doibase 10.1103/PhysRevD.77.107702} {\bibfield  {journal} {\bibinfo
  {journal} {Phys.Rev.}\ }\textbf {\bibinfo {volume} {D77}},\ \bibinfo {pages}
  {107702} (\bibinfo {year} {2008})},\ \Eprint {http://arxiv.org/abs/0803.1566}
  {arXiv:0803.1566 [hep-th]} \BibitemShut {NoStop}%
%%CITATION = ARXIV:0803.1566;%%
\bibitem [{\citenamefont {Arodz}\ and\ \citenamefont
  {Lis}(2009)}]{Arodz:2008nm}%
  \BibitemOpen
  \bibfield  {author} {\bibinfo {author} {\bibfnamefont {H.}~\bibnamefont
  {Arodz}}\ and\ \bibinfo {author} {\bibfnamefont {J.}~\bibnamefont {Lis}},\
  }\href {\doibase 10.1103/PhysRevD.79.045002} {\bibfield  {journal} {\bibinfo
  {journal} {Phys.Rev.}\ }\textbf {\bibinfo {volume} {D79}},\ \bibinfo {pages}
  {045002} (\bibinfo {year} {2009})},\ \Eprint {http://arxiv.org/abs/0812.3284}
  {arXiv:0812.3284 [hep-th]} \BibitemShut {NoStop}%
%%CITATION = ARXIV:0812.3284;%%
\bibitem [{\citenamefont {Arodz}\ \emph {et~al.}(2012)\citenamefont {Arodz},
  \citenamefont {Karkowski},\ and\ \citenamefont
  {Swierczynski}}]{Arodz:2012zh}%
  \BibitemOpen
  \bibfield  {author} {\bibinfo {author} {\bibfnamefont {H.}~\bibnamefont
  {Arodz}}, \bibinfo {author} {\bibfnamefont {J.}~\bibnamefont {Karkowski}}, \
  and\ \bibinfo {author} {\bibfnamefont {Z.}~\bibnamefont {Swierczynski}},\
  }\href {\doibase 10.5506/APhysPolB.43.79} {\bibfield  {journal} {\bibinfo
  {journal} {Acta Phys. Polon.}\ }\textbf {\bibinfo {volume} {B43}},\ \bibinfo
  {pages} {79} (\bibinfo {year} {2012})},\ \Eprint
  {http://arxiv.org/abs/1201.2279} {arXiv:1201.2279 [hep-th]} \BibitemShut
  {NoStop}%
%%CITATION = ARXIV:1201.2279;%%
\bibitem [{\citenamefont {Maldacena}(1999)}]{Maldacena:1997re}%
  \BibitemOpen
  \bibfield  {author} {\bibinfo {author} {\bibfnamefont {J.~M.}\ \bibnamefont
  {Maldacena}},\ }\href {\doibase 10.1023/A:1026654312961} {\bibfield
  {journal} {\bibinfo  {journal} {Int. J. Theor. Phys.}\ }\textbf {\bibinfo
  {volume} {38}},\ \bibinfo {pages} {1113} (\bibinfo {year} {1999})},\ \bibinfo
  {note} {[Adv. Theor. Math. Phys.2,231(1998)]},\ \Eprint
  {http://arxiv.org/abs/hep-th/9711200} {arXiv:hep-th/9711200 [hep-th]}
  \BibitemShut {NoStop}%
%%CITATION = HEP-TH/9711200;%%
\bibitem [{\citenamefont {Witten}(1998)}]{Witten:1998qj}%
  \BibitemOpen
  \bibfield  {author} {\bibinfo {author} {\bibfnamefont {E.}~\bibnamefont
  {Witten}},\ }\href@noop {} {\bibfield  {journal} {\bibinfo  {journal} {Adv.
  Theor. Math. Phys.}\ }\textbf {\bibinfo {volume} {2}},\ \bibinfo {pages}
  {253} (\bibinfo {year} {1998})},\ \Eprint
  {http://arxiv.org/abs/hep-th/9802150} {arXiv:hep-th/9802150 [hep-th]}
  \BibitemShut {NoStop}%
%%CITATION = HEP-TH/9802150;%%
\bibitem [{\citenamefont {Brodsky}\ \emph {et~al.}(2012)\citenamefont
  {Brodsky}, \citenamefont {Cao},\ and\ \citenamefont
  {de~Teramond}}]{Brodsky:2011sk}%
  \BibitemOpen
  \bibfield  {author} {\bibinfo {author} {\bibfnamefont {S.~J.}\ \bibnamefont
  {Brodsky}}, \bibinfo {author} {\bibfnamefont {F.-G.}\ \bibnamefont {Cao}}, \
  and\ \bibinfo {author} {\bibfnamefont {G.~F.}\ \bibnamefont {de~Teramond}},\
  }\href {\doibase 10.1088/0253-6102/57/4/21} {\bibfield  {journal} {\bibinfo
  {journal} {Commun. Theor. Phys.}\ }\textbf {\bibinfo {volume} {57}},\
  \bibinfo {pages} {641} (\bibinfo {year} {2012})},\ \Eprint
  {http://arxiv.org/abs/1108.5718} {arXiv:1108.5718 [hep-ph]} \BibitemShut
  {NoStop}%
%%CITATION = ARXIV:1108.5718;%%
\end{thebibliography}%
\end{document}